\documentclass[aps,prb,twocolumn,showpacs,superscriptaddress,groupedaddress,nofootinbib]{revtex4}  
\usepackage{mathbbold}
\usepackage{bm}
\usepackage{mathrsfs}
\usepackage{amsmath}
\usepackage{amssymb}
\usepackage{graphicx}
\usepackage{amsfonts}
\usepackage{cancel}
\usepackage{amsthm}
\usepackage{color}
\usepackage{dcolumn}
\usepackage{txfonts}
\usepackage{subfigure}
\begin{document}



\title{Nonlinear optical response induced by non-Abelian Berry curvature in time-reversal-invariant insulators}

\author{ Fan Yang }
\affiliation{Department of Physics, The Chinese University of Hong Kong, Shatin, N.T., Hong Kong, China}

\author{ Ren-Bao Liu }
\email{rbliu@phy.cuhk.edu.hk}
\affiliation{Department of Physics, The Chinese University of Hong Kong, Shatin, N.T., Hong Kong, China}
\affiliation{Centre for Quantum Coherence, The Chinese University of Hong Kong, Shatin, N.T., Hong Kong, China}
\affiliation{Institute of Theoretical Physics, The Chinese University of Hong Kong, Shatin, N.T., Hong Kong, China}



\begin{abstract}
We propose a general framework of nonlinear optics induced by non-Abelian Berry curvature in time-reversal-invariant (TRI) insulators. We find that the third-order response of a TRI insulator under optical and terahertz light fields is directly related to the integration of the non-Abelian Berry curvature over the Brillouin zone. We apply the result to insulators with rotational symmetry near the band edge. Under resonant excitations, the optical susceptibility is proportional to the flux of the Berry curvature through the iso-energy surface, which is equal to the Chern number of the surface times $2\pi$. For the III-V compound semiconductors, microscopic calculations based on the six-band model give a third-order susceptibility with the Chern number of the iso-energy surface equal to three.
\end{abstract}

\maketitle

When the parameters of a system are adiabatically varied, a non-degenerate eigenstate can acquire a Berry phase in addition to the dynamical one.~\cite{Berry} The Berry phase can be expressed as the flux of a gauge invariant field---Berry curvature---in the parameter space.~\cite{Berry,phase} This formalism can be extended to the cases with multiple (near-)degenerate energy bands, where the Berry phase and Berry curvature become matrices with non-Abelian gauge structures.~\cite{QNiu2} Berry curvatures play important roles in many fields of condensed matter physics, such as the quantum Hall effect,~\cite{Avron,TKNN,QNiu,Prange,QNiu2} the anomalous Hall effect,~\cite{QNiu2,Haldane_anomal,AH1,AH2,AH3,AH4} and topological insulators (TIs).~\cite{SCZhang,Qi} In quantum Hall effect, the Hall conductivity is proportional to the integration of the Berry curvature over the Brillouin zone, which is quantized by the Chern number and reflects the topology of the system.~\cite{TKNN} The anomalous Hall effect can be regarded as an ``unquantized" quantum Hall effect, where the integration of Berry curvature over part of the Brillouin zone gives rise to an intrinsic anomalous Hall conductivity.~\cite{QNiu2,Haldane_anomal} However, the Hall experiment cannot be used to probe the Berry curvature in TRI insulators (e.g. topological insulators), since the Hall conductivity is always equal to zero as protected by the time-reversal (TR) symmetry.




Berry curvatures appear naturally in various optical effects of condensed matter systems as revealed by recent works. For example, the Faraday rotation of a terahertz (THz) light in a quantum Hall system is found to be proportional to the Hall conductivity of the state, which presents plateau structures.~\cite{Optical_Hall1,Optical_Hall2} In recent studies on the interaction between polarized light beams and TR invariant spin currents, the Berry curvature dependence of the optical susceptibilities was noticed.~\cite{spin_current1st,spin_current} Hosur studied the optical response of the surface states of a three-dimensional TI and identified a Berry curvature dependent photogalvanic effect.~\cite{CPGE} Virk and Sipe discovered that a circularly polarized optical pulse can induce a transient macroscopic Berry curvature in semiconductors, which can be further detected by a THz light.~\cite{Sipe} Ref. \citenum{YF_MoS2} and \citenum{YF_bilayer} demonstrate the Berry phase effects in extreme nonlinear optics of semiconductors under strong THz fields, where the Berry phase of the quantum trajectory of an electron-hole pair induces a Faraday rotation of the optical emission. The Berry curvature effects in these works have essentially the same origin, that is, the matrix elements of the polarization operator in the basis of Bloch states are explicitly related to the Berry connection:~\cite{Blount}
\begin{equation}\label{dipole_moment}
\left\langle {\psi _{n,\mathbf{p}} } \right|\mathbf{r}\left| {\psi _{m,\mathbf{p}'} } \right\rangle  = i\left[ {\delta_{nm} \nabla _\mathbf{p}  + \left\langle {u_{n,\mathbf{p}} } \right|\nabla _\mathbf{p} \left| {u _{m,\mathbf{p}} } \right\rangle } \right]\delta \left( {\mathbf{p} - \mathbf{p}'} \right),
\end{equation}
where $\left| {\psi _{n,\mathbf{p}} } \right\rangle = e^{i\mathbf{p} \cdot \mathbf{r}} \left| {u _{n,\mathbf{p}} } \right\rangle$ is the Bloch state of the energy band.

We expect that the nonlinear optical spectroscopy, which is more flexible than transport approaches, can be used to directly detect the Berry curvature of TRI insulators. Since the second-order optical response vanishes in systems with inversion symmetry, we consider the next order nonlinear response, i.e. the third-order optical response. In general, the Berry connections of the conduction and valence bands are non-Abelian, i.e. the off-diagonal terms of the Berry connection are nonzero and thus can induce transitions between degenerate states according to equation (\ref{dipole_moment}). Through microscopic calculations, we find that the third-order response of a TRI insulator is nonzero and proportional to the integration of the non-Abelian Berry curvature. The basic physics of our work shares similarity with that of Ref.~\citenum{Sipe}. However, the theory in Ref.~\citenum{Sipe} is essentially an Abelian one. Furthermore, the transient THz response in Ref.~\citenum{Sipe} depends greatly on the shape of the optical injection pulse and the electron/hole decoherence. On the other hand, the steady nonlinear optical method in this paper would allow us to explore some interesting Berry curvature induced optical effects. For example, we find that if the system is (approximately) rotationally symmetric near the symmetry point, the third-order optical response under resonant excitations is quantized by the Berry curvature flux through the iso-energy surface, i.e. the Chern number of the surface. We apply the method to the six-band model of III-V compound semiconductors,~\cite{eightband_model} and obtain a quantized third-order susceptibility with Chern number equal to three.

\begin{figure}
\includegraphics[height=9cm]{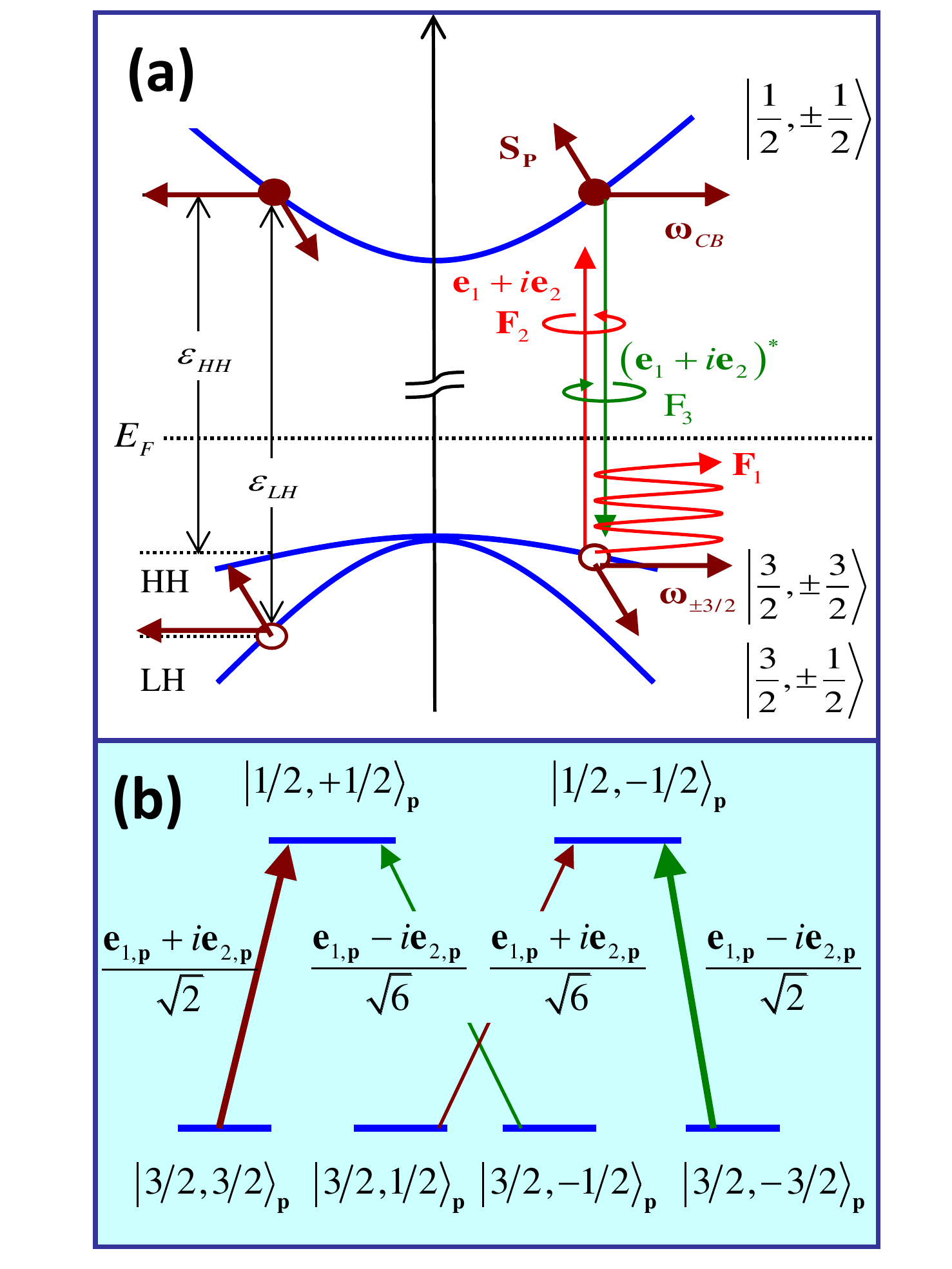}
\caption{ (a) Six-band model for the third-order optical response. The split-off bands (SO) are neglected since we only consider the near resonant transitions between the CB and LH/HH bands. We set up a coordinate system $(\mathbf{e}_{1,\mathbf{p}},\mathbf{e}_{2,\mathbf{p}},\mathbf{e}_{3,\mathbf{p}})$ at each momentum $\mathbf{p}$ so that $\mathbf{e}_{3,\mathbf{p}}=\hat p=\mathbf{p}/p$. $\mathbf{e}_{3,\mathbf{p}}$ is also the quantization direction of the spin states. $\mathbf{S}_{\mathbf{p}}$ denotes the spin polarization and $\bm{\omega}_{CB/ \pm 3/2}$ is the Berry curvature vector. (b) Selection rules and relative interband dipole moments from HH and LH bands to the CB.~\cite{spin_current}} \label{System}
\end{figure}

To understand the microscopic mechanism of the third-order optical effect, we consider the six-band model shown in Fig.\ref{System} that has both TR and inversion symmetry. Since the Fermi surface lies in the energy gap, no electron or hole exists in the system initially. If we use a linear-polarized THz light field, $\mathbf{F}_{1}$, to probe the system, the Faraday rotation vanishes due to the TR symmetry. Nevertheless, when two circularly polarized near resonant optical fields, $\mathbf{F}_2$ and $\mathbf{F}_3$, are applied to the system, a net spin polarization will form. Because the interband transitions are both spin and momentum dependent, the spin polarization breaks both the TR and inversion symmetry. For the sake of simplicity, we first consider the heavy-hole (HH) and conduction bands (CB), but neglect the light-hole (LH) bands. The difference-frequency process shown in Fig.\ref{System} creates an imbalance between the hole populations of the $\pm 3/2$ states, which is proportional to ${\left| {d_{cv} } \right|^2 }\mathbf{F}_2  \cdot \left[ {\left( {{\bf{e}}_{1,\mathbf{p}}  - i{\bf{e}}_{2,\mathbf{p}} } \right)\left( {{\bf{e}}_{1,\mathbf{p}}  + i{\bf{e}}_{2,\mathbf{p}} } \right) - c.c.} \right] \cdot \mathbf{F}_3^* \propto \left(\mathbf{F}_2  \times \mathbf{F}_3^* \right) \cdot \left| {d_{cv} } \right|^2 \hat p$, where $\left| {d_{cv} } \right|^2 \hat p$ gives the spin polarization. Therefore the two lights play the role of an effective ``magnetic field" given by $\mathbf{F}_2 \times \mathbf{F}_3^*$.~\cite{Magneto_Optics} The spin-polarized electrons and holes cause a nonzero Faraday rotation of the THz field, $\mathbf{F}_1$, which is related to the optical Hall conductance and in turn the Berry curvature of the energy bands.~\cite{Optical_Hall1,Optical_Hall2}

Guided by this observation, we calculate the third-order optical response of frequency $\omega_1 + \omega_2 - \omega_3$ using standard perturbation theory.~\cite{YRShen} The free Hamiltonian is
\begin{equation}
H_0  = \int {d \mathbf{p}} \left( {E_{ + , \mathbf{p}} \hat e_{\mu ,\mathbf{p}}^\dag  \hat e_{\mu ,\mathbf{p}}  - E_{ - ,\mathbf{p}} \hat h_{\mu , - \mathbf{p}}^\dag  \hat h_{\mu , - \mathbf{p}} } \right),
\end{equation}
where $\hat e$ and $\hat h$ are electron and hole operators, $+$ and $-$ are the indices of conduction and valence bands, respectively, $\mu$ is the (pseudo)spin index of the degenerate bands and summation of repeated dummy indices is assumed. We apply a THz probe light of frequency $\omega_1$ and two near resonant lights of frequencies $\omega_2$ and $\omega_3$ to the system. The polarization operator in terms of the Bloch states of the crystal is
\begin{equation}
\hat {\mathbf{P}} = \hat {\mathscr{P}} + \hat {\mathscr{D}}^\dag   + \hat {\mathscr{D}},
\end{equation}
where $\hat {\mathscr{P}}$ is the intraband polarization operator
\begin{equation}
\hat {\mathscr{P}} =  - ie\int {d \mathbf{p}} \left( {\hat e_{\mu ,\mathbf{p}}^\dag  {\mathbf{D}}_{\mu \nu,\mathbf{p}}^ +  \hat e_{\nu ,\mathbf{p}}  + \hat h_{\mu , - \mathbf{p}} {\mathbf{D}}_{\mu \nu,\mathbf{p}}^ -  \hat h_{\nu , - \mathbf{p}}^\dag  } \right),
\end{equation}
with $ {\mathbf{D}}_{\mu \nu,\mathbf{p}}^ \pm   = \delta _{\mu \nu } \nabla _\mathbf{p}  + \left\langle {\mu , \pm , \mathbf{p}} \right|\nabla _\mathbf{p} \left| {\nu , \pm , \mathbf{p}} \right\rangle$ being the covariant derivative. $\hat {\mathscr{D}}+\hat {\mathscr{D}}^\dag$ is the interband polarization operator with
\begin{equation}
\hat {\mathscr{D}}^\dag  = \int {d \mathbf{p}} \hat e_{\mu ,\mathbf{p}}^\dag  \hat h_{\nu , - \mathbf{p}}^\dag  \mathbf{d}_{\mu \nu,\mathbf{p}} ,
\end{equation}
where $ {\mathbf{d}}_{\mu \nu,\mathbf{p}}  =  - ie \left \langle {\mu , + ,\mathbf{p}} \right|\nabla _\mathbf{p} \left| {\nu , - , \mathbf{p}} \right\rangle$ is the interband dipole moment.~\cite{Blount} Thus, the interaction Hamiltonian under the rotating wave approximation is
\begin{equation}
\hat H_I = -\hat {\mathscr{P}} \cdot \mathbf{F}_1  e^{ - i\omega _1 t}  - \hat {\mathscr{D}}^\dag   \cdot \mathbf{F}_2  e^{ - i\omega _2 t}  - \hat {\mathscr{D}} \cdot \mathbf{F}_3^* e^{i\omega _3 t}  + h.c.
\end{equation}
In the interaction picture the third-order response is
\begin{align}
\notag \mathbf{P}^{\left( 3 \right)} \left( t \right) = & i \int\limits_{ - \infty }^t {d\tau } \int\limits_{ - \infty }^\tau  {dt'} \int\limits_{ - \infty }^{t'} {dt''}\\
&{\rm{Tr}}\left[ { \tilde {\mathscr{P}}\left( t \right)\left[ {\tilde H_I \left( \tau  \right),\left[ { \tilde H_I \left( {t'} \right),\left[ { \tilde H_I \left( {t''} \right),\hat \rho _0 } \right]} \right]} \right]} \right],
\end{align}
where $\hat \rho _0$ is the equilibrium density matrix of the ``vacuum" state with $\hat e_{\mu ,\mathbf{p}} \hat \rho _0  = 0$ and $\hat h_{\mu , - \mathbf{p}} \hat \rho _0  = 0$. We focus on the induced Faraday rotation of the THz field, $\mathbf{F}_{1}$. The corresponding susceptibility is
\begin{equation}
\tilde {\chi} _{ij,FR} = \frac{1}{2}\left( {\tilde \chi _{ij} -  \tilde \chi _{ji}} \right),
\end{equation}
where $\tilde {\bm \chi}  = \bm \chi ^{\left( 3 \right)} : \mathbf{F}^* _3 \mathbf{F}_2$ and $\bm \chi ^{\left( 3 \right)}$ is the third-order susceptibility defined through $\mathbf{P}^{\left( 3 \right)} \left( \omega = \omega_1+\omega_2-\omega_3 \right) = \bm \chi ^{\left( 3 \right)} \vdots \mathbf{F}^* _3 \mathbf{F}_2 \mathbf{F}_1$.

After a lengthy but straightforward calculation (details can be found in the supplementary information online), we obtain the susceptibility $\tilde \chi _{ij,FR} =\tilde \chi^A _{ij,FR}+ \tilde \chi^B _{ij,FR}$, where the Berry curvature dependent part is
\begin{align}
&\notag \tilde \chi^B _{ij,FR} \left( \omega \right)= \int \frac{-e^2d\mathbf{p}}{2\left( {2\pi } \right)^{d}} \left( {\frac{1}{{\omega _1  - \omega _3  + \varepsilon _\mathbf{p} }} + \frac{1}{{\omega _1  + \omega _2  - \varepsilon _\mathbf{p} }}} \right) \times \\
&\frac{{\mathbf{F}_2  \cdot {\mathbf{d}}_{\mu \alpha ,\mathbf{p}} {\mathbf{d}}_{\alpha \nu ,\mathbf{p}}^\dag  \cdot \mathbf{F}_3^* \left( {\Omega _{i j }^ +  } \right)_{\nu \mu }  - \mathbf{F}_3^*  \cdot {\mathbf{d}}_{\mu \alpha ,\mathbf{p}}^\dag  {\mathbf{d}}_{\alpha \nu ,\mathbf{p}} \cdot \mathbf{F}_2 \left( {\Omega _{i j }^ -  } \right)_{\nu \mu } }}{{\left( {\omega _2  - \varepsilon _\mathbf{p} } \right)\left( { - \omega _3  + \varepsilon _\mathbf{p} } \right)}}. \label{susceptB}
\end{align}
and
\small
\begin{align}
&\notag \tilde \chi^A _{ij,FR} \left( \omega \right) =  \int \frac{-e^2d\mathbf{p}}{2\left( {2\pi } \right)^{d}\left( {\omega _1  + \omega _2  - \omega _3 } \right)} \varepsilon _{ijl} \\
&\notag \left\{ {\mathbf{F}_2  \cdot  {\mathbf{d}}_{\mu \alpha ,\mathbf{p}} {\mathbf{d}}_{\alpha \nu ,\mathbf{p}}^\dag  \cdot \mathbf{F}_3^* \left( {{\mathbf{v}}_\mathbf{p}  \times \mathbf{A}_{\nu \mu ,\mathbf{p}}^ +  } \right)_l  - \mathbf{F}_3^*  \cdot {\mathbf{d}}_{\mu \alpha ,\mathbf{p}}^\dag {\mathbf{d}}_{\alpha \nu ,\mathbf{p}} \cdot \mathbf{F}_2 \left( {{\mathbf{v}}_\mathbf{p}  \times \mathbf{A}_{\nu \mu ,\mathbf{p}}^ -  } \right)_l } \right.\\
&\notag \left. { + \delta_{\mu\nu}\left[\left( {{\mathbf{v}}_\mathbf{p}  \times \nabla _\mathbf{p} } \right)_l \left( {\mathbf{F}_2  \cdot {\mathbf{d}}_{\mu \alpha ,\mathbf{p}} } \right)\right]{\mathbf{d}}_{\alpha \nu ,\mathbf{p}}^\dag   \cdot \mathbf{F}_3^* } \right\} \times\\
&\notag \left[ {\frac{1}{{\left( {\omega _3  - \varepsilon _\mathbf{p} } \right)\left( {\omega _1  - \omega _3  + \varepsilon _\mathbf{p} } \right)}}\left( {\frac{1}{{\omega _1  + \omega _2  - \omega _3 }} + \frac{1}{{\omega _1  - \omega _3  + \varepsilon _\mathbf{p} }}} \right)} \right. \\
&\left. { + \frac{1}{{\left( {\omega _2  - \varepsilon _\mathbf{p} } \right)\left( {\omega _1  + \omega _2  - \varepsilon _\mathbf{p} } \right)}}\left( {\frac{1}{{\omega _1  + \omega _2  - \omega _3 }} + \frac{1}{{\omega _1  + \omega _2  - \varepsilon _\mathbf{p} }}} \right)} \right], \label{susceptA}
\end{align}
\normalsize
Here $d$ is the dimension of the system, $\varepsilon _\mathbf{p}=E_{ + ,\mathbf{p}}-E_{ - ,\mathbf{p}}$ is the energy of the electron-hole pair, ${\mathbf{v}}_\mathbf{p}=\nabla_\mathbf{p} \varepsilon _\mathbf{p}$ is the semiclassical velocity, ${ \mathbf{A}_{\mu \nu ,\mathbf{p}}^ {\pm}   = \left\langle {\mu , \pm ,\mathbf{p}} \right|\nabla _\mathbf{p} \left| {\nu , \pm ,\mathbf{p}} \right\rangle }$ is the Berry connection and
\begin{equation}
\left( {\Omega _{i j }^ \pm  } \right)_{\nu \mu }  = \left[ {D_{p_i }^ \pm  ,D_{p_j }^ \pm  } \right]_{\nu \mu } = \partial _{p_i }  \mathbf{A}_{\nu \mu, p_j}^ {\pm}  + \sum\limits_{\kappa} \mathbf{A}_{\nu \kappa, p_i}^ {\pm} \mathbf{A}_{\kappa \mu, p_j}^ {\pm} - (i \leftrightarrow j)
\end{equation}
is the non-Abelian Berry curvature. In order to describe the dissipation in real materials, we can include a phenomenological dephasing $i\gamma$ in Eq. (\ref{susceptB}) and (\ref{susceptA}), i.e., $\omega _{2} \to \omega _{2} + i\gamma$ and $\omega _3 \to \omega _3 - i\gamma$. When the energy band is nearly flat, i.e. $v_\mathbf{p}=\left|\nabla_\mathbf{p} \varepsilon _\mathbf{p}\right|$ is sufficiently small, the response is dominated by the Berry curvature dependent part (\ref{susceptB}). However, unlike the quantum Hall effect, the nonlinear optical susceptibility is generally not quantized by the Chern number since the interband dipole moments vary greatly with the momentum $\mathbf p$ over the Brillouin-zone.

\begin{figure}
\begin{center}
\includegraphics[width=8cm]{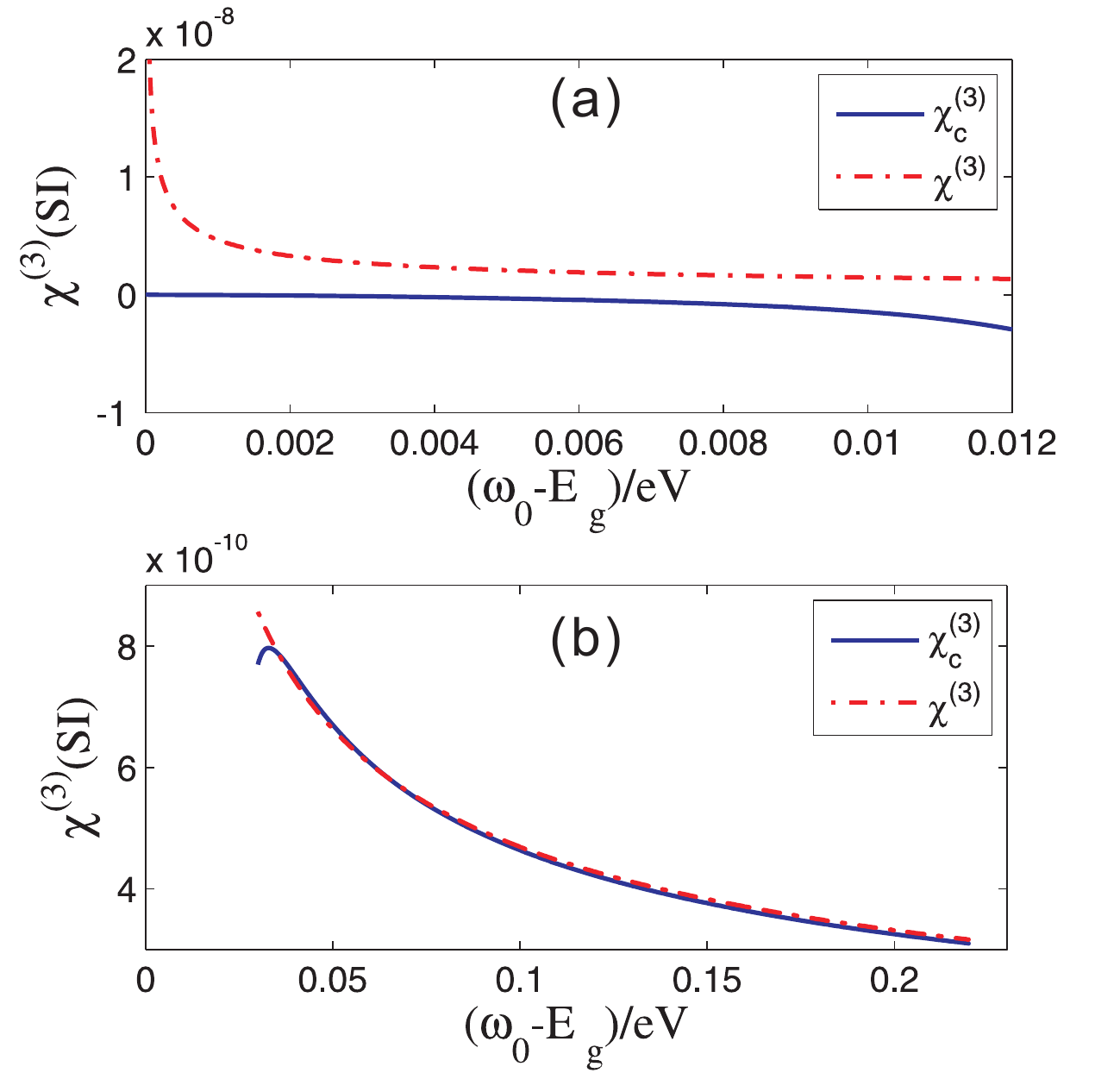}
\end{center}
\caption{Third-order susceptibility of GaAs with $\omega_1=8meV$, $\omega_2-\omega_3=0.15meV$ and the energy band broadening $\gamma=0.05meV$. The effective mass of the CB electron, HH and LH is in turn 0.067, 0.45 and 0.082 in units of the free electron mass, the dielectric constant $\varepsilon_r=10.6$ and the dipole moment $d_{cv}=6.7e\AA$. The solid lines show the complete susceptibilities calculated using the six-band model and the dashed lines show the susceptibilities given by (\ref{eightchi}). (a) The susceptibility when $\omega_0-E_g$ is very small, where $E_g$ is the band gap. Eq. (\ref{eightchi}) is divergent at the $\Gamma$ point due to the singular monopole, while the complete susceptibility is nearly zero since the LH and HH bands merge into the single four-band with topological charge equal to 0. (b) The susceptibility when $\omega_0-E_g$ is large, where the formula (\ref{eightchi}) is a good approximation.}
\label{eight}
\end{figure}

We consider another case where the laser frequencies, $\omega_{2}$ and $\omega_{3}$, are resonant with the transition energy $\varepsilon_{\mathbf{p}}$ (i.e. real absorption occurs in the system), $ \left| {\omega _2  - \omega _3 } \right| \ll \omega _1  \ll E_g$ and $\mathbf{F}_2  \parallel \mathbf{F}_3$. Then the real part of the signal is dominated by the Berry curvature dependent term $\tilde \chi^B _{ij,FR}$ (see the supplementary information online):
\begin{eqnarray}
&&\notag {\mathop{\rm Re}\nolimits} \left[ \tilde \chi _{ij,FR}  \right]
\approx \int \frac{ie^2 d \mathbf{p}}{(2\pi)^{d-1}} \frac{\delta \left( {\omega _0  - \varepsilon _\mathbf{p} } \right)}{\left( {\omega _2  - \omega _3 } \right)\left( {\omega _1  + \omega _2  - \omega _3 } \right)}\times \\
&&  \left[\mathbf{F}_2  \cdot \mathbf{d}_{\mu \alpha ,\mathbf{p}} \mathbf{d}_{\alpha \nu ,\mathbf{p}}^\dag \cdot \mathbf{F}_3^* \left( {\Omega _{i j }^ +  } \right)_{\nu \mu }  - \mathbf{F}_3^*  \cdot \mathbf{d}_{\mu \alpha ,\mathbf{p}}^\dag  \mathbf{d}_{\alpha \nu ,\mathbf{p}} \cdot \mathbf{F}_2 \left( {\Omega _{i j }^ -  } \right)_{\nu \mu } \right],\notag \\ \label{resonant}
\end{eqnarray}
where $\omega_0=(\omega_2+\omega_3)/2$.
From formula (\ref{resonant}), we see that only the electron-hole pairs with energy close to $\omega_0$ can be excited and the real part of the susceptibility is reduced to an integral of the Berry curvature over a two-dimensional iso-energy surface. In insulators with both TR and inversion symmetry, the conduction and valence bands have two-fold degeneracy, which is denoted by the pseudospin index $\mu=\Uparrow/\Downarrow$. Also we have
\begin{subequations}
\begin{equation}
{\mathbf{d}}_{\Downarrow \alpha ,\mathbf{p}} {\mathbf{d}}_{\alpha \Downarrow,\mathbf{p}}^\dag = \left( {\mathbf{d}}_{\Uparrow \alpha ,-\mathbf{p}}  {\mathbf{d}}_{\alpha \Uparrow,-\mathbf{p}}^\dag   \right)^*  = \left( {\mathbf{d}}_{\Uparrow \alpha ,\mathbf{p}}  {\mathbf{d}}_{\alpha \Uparrow,\mathbf{p}}^\dag   \right)^*,
\end{equation}
\begin{equation}
\left( {\Omega _{i j }^ {\pm} \left(\mathbf p\right)} \right)_{\Downarrow \Downarrow} = \left( {\Omega _{i j }^ {\pm} \left(-\mathbf p\right)} \right)^*_{\Uparrow \Uparrow}  = - \left( {\Omega _{i j }^ {\pm} \left(\mathbf p\right)} \right)_{\Uparrow \Uparrow}.
\end{equation}
\end{subequations}
Although the Berry connection is generally non-Abelian, the off-diagonal terms of the Berry curvature must vanish since
\begin{equation}
\left( {\Omega _{i j }^ {\pm} \left(\mathbf p\right)} \right)_{\Uparrow \Downarrow} = \left( {\Omega _{i j }^ {\pm} \left(-\mathbf p\right)} \right)^*_{\Downarrow \Uparrow} = -\left( {\Omega _{i j }^ {\pm} \left(-\mathbf p\right)} \right)_{\Uparrow \Downarrow} = - \left( {\Omega _{i j }^ {\pm} \left(\mathbf p\right)} \right)_{\Uparrow \Downarrow}.
\end{equation}
Then the susceptibility (\ref{resonant}) is reduced to
\begin{equation}\label{resonant2}
{\mathop{\rm Re}\nolimits} \left[ {\tilde \chi _{ij,FR} } \right]
= \int\limits_{\varepsilon _\mathbf{p}  = \omega _0 } {\frac{{-2e^2 dS_\mathbf{p} }}{{(2\pi)^{d-1}v_\mathbf{p} }} \frac{\varepsilon _{ijl} \left( { \mathbf{F}_2  \times \mathbf{F}_3^* } \right) \cdot  \left( {\mathbf{s}_{\Uparrow,\mathbf{p}}^{+} \omega _{\Uparrow,l }^{+}  + \mathbf{s}_{\Uparrow, \mathbf{p}}^{ - } \omega _{\Uparrow,l }^{ - } } \right) }{ \left( {\omega _2  - \omega _3 } \right)\left( {\omega _1  + \omega _2  - \omega _3 } \right)}},
\end{equation}
where $S_\mathbf{p}$ is the iso-energy surface with $\varepsilon _\mathbf{p}  = \omega _0$. In Eq. (\ref{resonant2}), we have defined three vectors from the anti-symmetric tensors:
\begin{equation}
s_{\Uparrow,p_l }^{ + } = \frac{\varepsilon _{ijl}}{2} {\mathop{\rm Im}\nolimits} \left(  {\mathbf{d}}_{\Uparrow \alpha ,p_i } {\mathbf{d}}_{\alpha \Uparrow ,p_j }^\dag   \right) =\varepsilon _{ijl} \frac{  {\mathbf{d}}_{\Uparrow \alpha ,p_i } {\mathbf{d}}_{\alpha \Uparrow ,p_j }^\dag - {\mathbf{d}}_{\Uparrow \alpha ,p_j } {\mathbf{d}}_{\alpha \Uparrow ,p_i }^\dag  }{4i},
\end{equation}
and
\begin{equation}
s_{\Uparrow, p_l }^{ - }  = \frac{\varepsilon _{ijl}}{2} {\mathop{\rm Im}\nolimits} \left( {\mathbf{d}}_{\Uparrow \alpha ,p_i }^\dag  {\mathbf{d}}_{\alpha \Uparrow ,p_j } \right), \ \ \omega _{\Uparrow,l }^{ \pm }  = \frac{\varepsilon _{ijl}}{2} \left( {\Omega _{i j }^ \pm  } \right)_{\Uparrow \Uparrow}.
\end{equation}
Here $\mathbf{s}_{\Uparrow,\mathbf{p}}^{ +/- }$ is the spin polarization in the conduction/valence bands and $\mathbf{F}_2 \times \mathbf{F}_3^*$ is the effective ``magnetic field".~\cite{Magneto_Optics} Note that $\mathbf F_2 \times \mathbf F_3^*$ is not equal to zero although $\mathbf{F}_2  \parallel \mathbf{F}_3$, since $\mathbf F_2$ can be circularly polarized. Thus $ ( { \mathbf{F}_2  \times \mathbf{F}_3^* } ) \cdot   \mathbf{s}_{\Uparrow,\mathbf{p}}^{ \pm }$ gives the net spin population in the system created by the two near resonant lights, and $\varepsilon _{ijl} \omega _{\Uparrow,l }^{\pm}$ gives the Faraday rotation of the THz light caused by the spin-polarized particles. Since $1/v_{\mathbf{p}}$ is the density of states, $dS_\mathbf{p}/v_{\mathbf{p}}$ gives the number of states in the iso-energy surface element. Conceptually, the formula (\ref{resonant2}) describes a process that can be regarded as an optically-induced quantum Hall effect. If the insulator has rotational symmetry near the symmetry point, we have $\varepsilon_{\mathbf{p}}  = \varepsilon _p$, $\mathbf{s}_{\Uparrow,\mathbf{p}}^{  \pm }  = s^ \pm  \left( p \right)\hat {p}$ and
$\bm{\omega}_{\Uparrow,\mathbf{p}}^{ \pm }  = \omega^ \pm  \left( p \right)\hat {p}$. The third-order susceptibility then reduces to
\begin{equation}
{\mathop{\rm Re}\nolimits} \left[ {\tilde \chi _{ij,FR} } \right] = \frac{-2e^2 \varepsilon _{ijl} \left( {\mathbf{F}_2  \times \mathbf{F}_3^* } \right)_l}{{\left( {\omega _2  - \omega _3 } \right)\left( {\omega _1  + \omega _2  - \omega _3 } \right)}}  \left.{\frac{s^ + Q^+ + s^ - Q^-}{{3(2\pi)^{d-2}v_p }}}\right|_{\varepsilon _p  = \omega _0 }, \label{Final}
\end{equation}
where $Q^\pm=\int\limits_{\varepsilon _p  = \omega _0 } {\frac{dS_p}{2\pi} \omega ^ \pm  \left( p \right)}$ is the Berry curvature flux through
the iso-energy surface divided by $2\pi$, i.e. the Chern number of the surface. When there is a singular monopole (i.e. the band degeneracy point) in momentum space, the third-order susceptibility is proportional to its topological charge $Q^\pm$ and Eq. (\ref{Final}) becomes divergent at the singular point.



As an example, we apply the formula (\ref{Final}) to the III-V compound semiconductors. We assume that $\omega_0$ are tuned near the band edge. Thus we can neglect the SO bands and consider only the optical transitions between CB, HH and LH bands. If $\omega_0$ is tuned to the $\Gamma$ point (or a small region around it), the HH-LH splitting is negligible and the four hole bands constitute a basis of the irreducible representation with total angular momentum $3/2$.
In this case, the non-Abelian Berry curvatures of the CB and the hole bands both vanish. Thus, we obtain $\notag {\mathop{\rm Re}\nolimits} \left[ \tilde \chi _{ij,FR} \right] \approx 0$ (see Fig.\ref{eight}(a)) if we apply the formula (\ref{Final}) to the optical transitions between the CB and the four hole bands.
On the other hand, if $\omega_0$ is tuned to the region far apart from the $\Gamma$ point, the HH-LH splitting is much larger than $\omega_1$ such that the HH-LH transition caused by the THz light can be neglected. Then applying the formula (\ref{Final}) to the CB-HH and CB-LH separately, we obtain the susceptibility as
\begin{align}
\notag {\mathop{\rm Re}\nolimits} \left[ \tilde \chi _{ij,FR} \right]
= & \frac{i\left| {d_{cv} } \right|^2 e^2 \varepsilon _{ijl} \left( {\mathbf{F}_2  \times \mathbf{F}_3^* } \right)_l}{{6\pi \left( {\omega _2  - \omega _3 } \right)\left( {\omega _1  + \omega _2  - \omega _3 } \right)}} \times \\
&  \left[ {\left( {\frac{Q}{{3v_p^{LH} }}} \right)_{\varepsilon _p^{LH}  = \omega _0 }  - \left( {\frac{Q}{{v_p^{HH} }}} \right)_{\varepsilon _p^{HH}  = \omega _0 } } \right],\label{eightchi}
\end{align}
using $\mathbf{s}_{\mathbf{p}, \pm 1/2}^{LH}  =  \pm \frac{{\left| {d_{cv} } \right|^2 }}{6}\hat p$, $\bm{\omega} _{\mathbf{p}, \pm 1/2}^{LH}  =  \mp \frac{{3i}}{{2p^2 }}\hat p$, $\mathbf{s}_{\mathbf{p},\pm 3/2}^{HH}  =  \pm \frac{{\left| {d_{cv} } \right|^2 }}{2}\hat p$, $\bm{\omega} _{\mathbf{p}, \pm 3/2}^{HH}  =  \pm \frac{{3i}}{{2p^2 }}\hat p$ and $\bm {\omega} _{\mathbf{p}, \pm 1/2}^{CB}  = 0$. This is a quantized response with topological charge $Q=3$, although no plateau structure appears due to the variation of the density of states with $\varepsilon_p$. As an example, we apply the formula (\ref{eightchi}) to GaAs. In Fig.\ref{eight} we plot the figures of the macroscopic susceptibility $\chi^{\left(3\right)}$ defined from
\begin{equation}
\left(\frac{\varepsilon_r+2}{3}\right)^4\left(\mathbf{F}_{1}\right)_j \mathbf{e}_i {\mathop{\rm Re}\nolimits} \left[ {\tilde \chi _{ij,FR} } \right]=\varepsilon_0 i \left( {\mathbf{F}_2  \times \mathbf{F}_3^* } \right) \times \mathbf{F}_1 \chi^{\left(3\right)},
\end{equation}
and compare it with the complete susceptibility $\chi^{\left(3\right)}_c$ derived using the six-band model with the HH-LH transition included. The factor $(\varepsilon_r+2)/3$ takes into account the difference between the macroscopic external field and the microscopic local field.~\cite{Bloem} It can be seen that if $\omega_0-E_g$ is large enough such that the HH-LH splitting is much larger than $\omega_1$, the formula (\ref{eightchi}) is a good approximation [see Fig.\ref{eight}(b)].

In summary, with a general microscopic calculation, we found that the third-order susceptibility of insulators under optical and THz lights depends on the non-Abelian Berry curvature directly. For an insulator with rotational symmetry near the band edge, the resonant susceptibility is proportional to the topological charge of the energy band. This nonlinear spectroscopy method provides a new way to explore the Berry curvature as a fundamental property of insulators. Owing to the generality and controllability of the nonlinear spectroscopy technique, it can be used to study a wide variety of materials with nontrivial Berry curvatures such as topological insulators.~\cite{Qi,SCZhang,Kane,TI1,TI2,TI3,TI4,TI5} A wealth of new physics connecting nonlinear optics and topological properties of insulators may be discovered.

This work is supported by Hong Kong RGC/GRF CUHK401011 and CUHK Focused Investments Scheme.

\newcommand{\noopsort}[1]{} \newcommand{\printfirst}[2]{#1}
  \newcommand{\singleletter}[1]{#1} \newcommand{\switchargs}[2]{#2#1}

\end{document}